\documentclass[10pt]{article}
\pdfoutput=1

\usepackage{geometry}
\usepackage{graphicx} 

\usepackage{amsmath,amssymb}
\usepackage{hyperref}
\usepackage{cite}
\usepackage[utf8]{inputenc}
\usepackage[justification=justified,font=small,labelfont=bf]{caption}

\usepackage{tikz}
\usetikzlibrary{arrows,positioning,decorations.markings,decorations.pathmorphing,calc}

\newcommand{\ie}{\textit{i.e.}\ }
\newcommand{\eg}{\textit{e.g.}\ }
\newcommand{\etc}{\textit{etc.}}
\newcommand{\viz}{\textit{viz.}\ }

\title{Existence of Life in 2 + 1 Dimensions}
\author{J. H. C. Scargill\footnote{\tt
jhcscargill@gmail.com}\\ \\
{\it Center for Quantum Mathematics and Physics (QMAP), Department of Physics, }\\
{\it University of California at Davis, One Shields Avenue, Davis, CA 95616, USA}}
\date{}

\begin{document}
\maketitle


\begin{abstract}
There are anthropic reasons to suspect that life in more than three spatial dimensions is not possible, and if the same could be said of fewer than three, then one would have an anthropic argument for why we experience precisely three large spatial dimensions. 
There are two main arguments levelled against the possibility of life in $2+1$ dimensions: the lack of a local gravitational force and Newtonian limit in 3D general relativity, and the claim that the restriction to a planar topology means that the possibilities are ‘too simple’ for life to exist. I will examine these arguments and show how a purely scalar theory of gravity may evade the first one, before considering certain families of planar graphs which share properties which are observed in real-life biological neural networks and are argued to be important for their functioning.
\end{abstract}

\section{Introduction}

The question of why we experience the number of dimensions we do is one with a long history, which I do not attempt to cover here, instead simply referring the reader to \cite{Ehrenfest:1918, Whitrow:1955, Barrow:1983, Barrow:1988yia, Tegmark:1997jg} and references therein. Tegmark \cite{Tegmark:1997jg} convincingly argues for the existence of precisely one temporal dimension by requiring hyperbolic equations of motion, and hence predictability (which certainly does seem a requirement if scientists are to emerge), which leaves us with the question of why there are three (large) spatial dimensions. Various arguments for this have been proposed, including that it might be entropically or thermodynamically favoured \cite{Momen:2011jc, Gonzalez-Ayala:2015xda}, or that it might have a dynamical reason, such as string gas cosmology \cite{Brandenberger:1988aj, Greene:2012sa}, or evaporation of D-branes \cite{Durrer:2005nz}, or even that it might be due to properties of the Weyl equation \cite{Nielsen:1993fd}.

In the absence of a truly convincing argument, however, we may rely instead on anthropic reasoning. The anthropic argument against there being more than three large spatial dimensions is well known: in Newtonian gravity in more than three dimensions, orbits are unstable against small perturbations, and hence anything like a solar system is impossible.

The arguments against there being fewer than three seem, however, not so robust. A commonly cited one is the absence of propagating degrees of freedom in general relativity (GR) in $2+1$ dimensions. Were one forced to use purely GR, then this indeed would be a problem for the existence of solar systems, and thus life that is in any way similar to us; however, nothing forces the gravitational theory to be simply GR and, in particular, one could include a scalar field and thus reintroduce propagating, `gravitational' degrees of freedom, and hence dynamics that seem more amenable to the existence of complex life. In section \ref{sec-3D gravity} I will present a purely scalar theory of gravity which admits stable orbits and a possibly reasonable cosmology, as a proof-of-principle.

The other objection that often arises is that complex life would face topological obstructions in two dimensions; certainly one-way digestive systems would not be possible, though presumably life that evolved in such a world would not possess our squeamishness about this. On the other hand, neurons in the brain would not be able to cross, and hence the neural network of a brain would have to be planar, which, in some sense, limits its complexity. Thus \cite{Whitrow:1955, Tegmark:1997jg} conclude that a two dimensional world is too simple to allow complex life to appear. In the second half of the paper, section \ref{sec-neural networks}, I will analyse this question quantitatively, by examining various classes of planar networks, in order to determine whether it is possible for them to emulate some of the behaviour that has been observed in biological neural networks in the real world. 

Finally, in section \ref{sec-other things} I will briefly discuss the nature of electromagnetism and the existence and structure of atoms and molecules, both of which have a bearing on life.

One further objection that has been raised \cite{Ehrenfest:1918, Barrow:1983} is that a solution of the wave equation (for a massless field) in an even number of spatial dimensions in general has support on the interior of the past lightcone, not just on its surface. That is, massless fields transmit disturbances not just along the lightcone, which would seem to make the faithful transmission of information more difficult, and hence is an impediment to life's functioning. However, as the saying goes, ``life finds a way," and it is not inconceivable that beings which evolved under such circumstances would find a way to nonetheless communicate effectively. That being said, it is not clear to me how to test this supposition, and so for now I simply note it and move on.

Finally, it behoves me also to mention the pioneering work of A. K. Dewdney who has thought extensively about the possibility of life in two spatial dimensions and the form that it might take \cite{Dewdney:1979, 2dscitech, 2ndSymposiumon2dscitech}.\footnote{A more light-hearted presentation of these ideas may be found in \cite{Planiverse}.}

\section{Relativistic Gravity in Three Dimensions}\label{sec-3D gravity}

As is well known, general relativity in $2+1$ dimensions does not have any local degrees of freedom; as a result of this, the spacetime outside of, \eg , a star is locally flat, and the presence of the object is only discernible globally, through the presence of a deficit angle. Clearly solar systems could not exist in such a world, and so this is commonly used as an argument against the possibility of life in two spatial dimensions. The absence of local degrees of freedom in $2+1$ dimensions is however a peculiarity of the fact that GR comprises just a massless spin-2 degree of freedom,\footnote{In $D$ spacetime dimensions this comprises $\frac{1}{2}D(D-3)$ degrees of freedom.} and since we are already considering altering the number of spacetime dimensions, it seems not unreasonable to also modify the theory of gravity so as to include local degrees of freedom. The simplest way to do this is to include a gravitational scalar field, and for completeness and concreteness I will give an example of such a theory.

Consider the action
\begin{equation}
S = \mu \int d^3 x \sqrt{-\det (f(\phi)^2\eta_{\mu\nu}) } \left( - \frac{1}{2} k(\phi) f(\phi)^{-2} \eta^{\mu\nu} \partial_\mu \phi \partial_\nu \phi - V(\phi) \right) + S_\text{m}\left[ f(\phi)^2 \eta_{\mu\nu} , \{ \Psi \} \right],
\end{equation}
where $\mu$ is a mass parameter, and $\{ \Psi \}$ denotes the collection of matter fields. Note that this describes matter and a scalar field $\phi$ (possibly with a non-minimal kinetic term) minimally coupled to a conformally flat metric $g_{\mu\nu} = f(\phi)^2 \eta_{\mu\nu}$. For calculational ease, and before specifying the conformal coupling $f$, I will perform the field redefinition $d\varphi = \sqrt{f k} d\phi$ to get
\begin{equation}
S = \mu \int d^3x \left[ - \frac{1}{2} \eta^{\mu\nu} \partial_\mu \varphi \partial_\nu \varphi - U(\varphi) \right] + S_\text{m}\left[ f(\varphi)^2 \eta, \{ \Psi \} \right]. \label{scalar gravity action}
\end{equation}
When $f(\varphi) = \varphi$, this theory is essentially one possible generalisation to three dimensions of Nordstr\"om's theory of gravity.\footnote{In four dimensions one has $(\partial \varphi )^2 \sim \sqrt{-\det (\varphi ^2 \eta)} R(\varphi ^2 \eta)$ and so the action is the same as in GR, but with the restriction to conformally flat metrics; in three dimensions one has $\sqrt{-\det (\varphi ^2 \eta)} R(\varphi ^2 \eta) \sim \varphi^{-1} (\partial \varphi)^2$ from which one would get a different possible generalisation of Nordstr\"om's theory (of course, the kinetic term can be made canonical via a field redefinition, but doing so would change the form of the coupling to matter).} The field equation resulting from this action is
\begin{equation}
\Box \varphi - U' = - \mu^{-1}|f|' T, \label{scalar gravity field equation}
\end{equation}
where the traces are taken using the Minkowski metric, and the matter stress-energy tensor is defined with respect to the effective metric to which matter couples.\footnote{Note that one has $\frac{\delta S_\text{m}}{\delta \varphi} = \frac{\delta S_\text{m}}{\delta g_\text{eff}^{\mu\nu}} \frac{\delta g_\text{eff}^{\mu\nu}}{\delta \varphi} = - \frac{1}{2} \sqrt{- \det g_\text{eff}} T_{\mu\nu} \eta^{\mu\nu} \frac{d f^{-2}}{d \varphi} = |f|^3 \frac{f'}{f^3} T$.} I will take the scalar potential to be $U(\varphi) = \frac{1}{2}m^2 \left( \varphi + 1 \right)^2$, which is done in order to force the vev of $\varphi$ to be $\varphi = -1$ and, hence, in empty space $g_{\mu\nu} = \eta_{\mu\nu}$ when $f = \varphi$.\footnote{It should be noted that if $f = \varphi$ and $k = 1$ then $\varphi = \frac{1}{4}\phi^2$ and hence $V(\phi) = 2 m \phi^{-6} \left( \phi^2 + 4 \right)^2$.}

\subsection{Spherically symmetric, static source}

First let us consider the case relevant for planetary orbits: spherical symmetry, staticity, and $T_{\mu\nu} = M \delta(\mathbf{x}) \delta_\mu^0 \delta_\nu^0$,\footnote{Note that $T_{\mu\nu} = M \delta^{(g_\text{eff})}(\mathbf{x}) u_\mu u_\nu$, where $\delta^{(g_\text{eff})}(\mathbf{x}) = f^{-2} \delta(\mathbf{x})$ and $u_\mu = f \delta_\mu^0$.} for which the field equation becomes
\begin{equation}
\frac{1}{r} \frac{d}{d r} ( r \varphi') - m^2 \left( \varphi + 1 \right) = |f|' \frac{M}{\mu} \frac{\delta(r)}{\pi r}.
\end{equation}
The solution to the homogeneous equation is $\varphi = -1 + c_1 H_0^{(1)}(imr) + c_2 H_0^{(2)}(imr)$, where $H_0^{(1,2)}$ are Hankel functions of the first and second kind. Requiring $\varphi \to -1$ as $r \to \infty$ sets $c_2 = 0$, and $c_1$ is fixed by integrating over the singularity at $r = 0$ to get $4 i c_1 = \frac{M}{\mu} |f|' |_{r \to 0}$. In order for this to be finite, $f'$ must be finite, and the simplest solution is to choose $f = \varphi$, leading to
\begin{equation}
\varphi = -1 + \frac{M}{4 \mu} i H_0^{(1)}(imr),
\end{equation}
with matter feeling the metric
\begin{equation}
ds^2 = \left( 1 - \frac{M}{4 \mu} i H_0^{(1)}(imr) \right)^2 \left( -dt^2 + dr^2 + r^2 d\theta^2 \right).
\end{equation}

Let us define a new radial coordinate via $R = \left(1 - \frac{M}{4 \mu} i H_0^{(1)}(imr) \right) r$, with which the above metric can be written as
\begin{equation}
ds^2 = - \left( 1 - \frac{M}{4 \mu} i H_0^{(1)}(imr) \right)^2 dt^2 + \left( \frac{1 - \frac{M}{4 \mu} i H_0^{(1)}(imr)}{1 - \frac{M}{4 \mu}( i H_0^{(1)}(imr) + mr H_1^{(1)}(imr) )} \right)^2 dR^2 + R^2 d\theta^2. \label{spherically symmetric metric}
\end{equation}

The conserved energy of test particles moving in this metric is given by
\begin{equation}
E^2 = \frac{\left( 1 - \frac{M}{4 \mu} i H_0^{(1)}(imr) \right)^4}{ \left(1 - \frac{M}{4 \mu}( i H_0^{(1)}(imr) + mr H_1^{(1)}(imr) ) \right)^2} \dot{R}^2 + V(R),
\end{equation}
where the effective potential is given by
\begin{equation}
V(R) = \left( 1 - \frac{M}{4 \mu} i H_0^{(1)}(imr) \right)^2 + \frac{J^2}{r^2},
\end{equation}
where $J$ is the conserved angular momentum. The first of the terms in the effective potential increases monotonically from 0 to 1, while the second decreases monotonically from $J^2/r_0^2$ (where $R(r_0) = 0$) to 0, hence, for a small enough $J$, their sum will possess a minimum at a finite value of $R$, and thus the metric \eqref{spherically symmetric metric} supports \emph{stable} circular orbits.\footnote{Since the conformal factor is $f^2$, choosing the vev of $\varphi$ to be $+1$ would also be consistent with flat space at infinity, however in this case the matching condition for the coefficient of $H_0^{(1)}$ would have been unaffected and so the metric would have been $ds^2 = \left( 1 + \frac{M}{4 \mu} i H_0^{(1)}(imr) \right)^2 \left( -dt^2 + dr^2 + r^2 d\theta^2 \right)$. The corresponding sign is flipped in the first term of $V(R)$ and hence both terms are monotonically decreasing, disallowing the existence of orbits in this case.}

It behoves me to mention one undesirable aspect of this solution, which is that the metric \eqref{spherically symmetric metric} exhibits a naked singularity at $R = 0$ as can be seen by calculating the Ricci scalar.

\subsection{Homogeneous, isotropic universe}

Other solutions which are important for life are cosmological ones (see \cite{Barrow:2006cw} for a study of cosmological solutions in three-dimensional general relativity). To that end let us consider a homogeneous and isotropic universe filled with a perfect fluid:
\begin{equation}
T_{\mu\nu} = \varphi^2 \left( (\rho + p) \delta_\mu^0 \delta_\nu^0 + p \eta_{\mu\nu} \right).
\end{equation}
The field equation \eqref{scalar gravity field equation} gives
\begin{equation}
\ddot{\varphi} + m^2 (\varphi + 1) = \mathrm{sign}(\varphi) \varphi^2 \frac{2p - \rho}{\mu}. \label{cosmo field equation}
\end{equation}
We must supplement this with the equation for energy-momentum conservation, $\partial^\mu \left( T^{(\varphi)}_{\mu\nu} + T^{(\eta, \text{m})}_{\mu\nu} \right) = 0$, where $T^{(\varphi)}_{\mu\nu}$ is the energy-momentum tensor coming from the $\varphi$ kinetic term, while $T^{(\eta, \text{m})}_{\mu\nu}$ is the energy-momentum tensor for the matter sector defined with respect to the Minkowski metric
which satisfies $T^{(\eta, \text{m})}_{\mu\nu} = |\varphi| T_{\mu\nu}$. Hence energy-momentum conservation yields
\begin{equation}
\frac{d}{d t} \left( \frac{1}{2} \dot{\varphi}^2 + \frac{1}{2} m^2 (\varphi + 1)^2 + \frac{\rho}{\mu} |\varphi|^3 \right) = 0. \label{cosmo energy conservation}
\end{equation}
By combining this with \eqref{cosmo field equation} one obtains the expected
\begin{equation}
\dot{\rho} + 2 \frac{\dot{\varphi}}{\varphi} (\rho + p) = 0,
\end{equation}
and hence a fluid with barotropic equation of state $p = w \rho$ behaves as $\rho \propto |\varphi|^{-2(1+w)}$. 



The dynamics of the universe can be easily ascertained from the `Friedmann equation' \eqref{cosmo energy conservation}, which tells us that the scale factor just behaves as a particle rolling in an effective potential $V(\varphi) = \frac{1}{2} m^2 (\varphi + 1)^2 + \frac{\rho}{\mu} |\varphi|^3$. Since the point-source solution described in the previous subsection has $\varphi < 0$, here I will only focus on $\varphi < 0$.

First note that radiation ($w = \frac{1}{2}$) only changes the zero-point of the effective potential and hence has no effect on the dynamics. For universes which also contain non-relativistic matter (of co-moving density $\tilde{\rho} > 0$) and vacuum energy (of density $\rho_v$) the qualitative behaviour is as follows.
\begin{description}
\item[$\rho_v > 0$ and $\tilde{\rho} > \mu m^2$:] the universe expands and recollapses.
\item[$\rho_v > 0$ and $\tilde{\rho} < \mu m^2$:] the universe either expands and recollapses, or it may oscillate around a finite size.
\item[$\rho_v < 0$ and $\tilde{\rho} > \mu m^2$:] the universe either expands and recollapses, or it may transition from matter to vacuum energy domination, ending in a big-rip.
\item[$\rho_v < 0$ and $\tilde{\rho} < \mu m^2$:] the universe either expands and recollapses, or it may oscillate around a finite size, or it may transition from matter to vacuum energy domination, ending in a big-rip.
\end{description}


\subsubsection{Linear perturbations}

It is important to check the behaviour of linear perturbations in these universes. Perturbing the gravitational scalar and the non-relativistic matter, one arrives at the following equations for the perturbations
\begin{align}
\ddot{\delta \varphi} + (k^2 + m^2) \delta \varphi &= - \frac{1}{\mu} \delta \tilde{\rho}, \\
\varphi \ddot{\delta \tilde{\rho}} + \dot{\varphi} \dot{\delta \tilde{\rho}} &= - k^2 \tilde{\rho} \delta \varphi, \\
\dot{\delta u_i} &= i k_i \delta \varphi,
\end{align}
where $\tilde{\rho} = \rho \varphi^2$ is the (constant) co-moving matter density. A universe dominated by non-relativistic matter has
\begin{equation}
|\varphi(t)| = \frac{1}{2}\left( (\varphi_\text{max} + \varphi_\text{min}) + (\varphi_\text{max} - \varphi_\text{min}) \cos (mt) \right),
\end{equation}
where $\varphi_\text{max} = 2 \left(1 - \frac{\tilde{\rho}}{\mu m^2} \right) - \varphi_\text{min}$. Unfortunately, we cannot solve these exactly, but if one can drop the $\ddot{\delta \varphi}$ term in the first equation (a quasi-static approximation), then the second equation becomes
\begin{equation}
\delta \tilde{\rho}''(\tau) = \frac{k^2}{k^2 + m^2} \frac{\tilde{\rho}}{\mu m^2} \varphi(\tau) \delta \tilde{\rho}(\tau), \label{tau perturbation}
\end{equation}
where the new time coordinate is
\begin{equation}
dt = \varphi d \tau \quad \implies \quad \tau = \frac{2}{\sqrt{\varphi_\text{min} \varphi_\text{max}}} \tan^{-1} \left( \sqrt{\frac{\varphi_\text{min}}{\varphi_\text{max}}} \tan \left( \frac{m t}{2} \right) \right). \label{tau t}
\end{equation}

Equation \eqref{tau perturbation} clearly possesses a growing solution, and so we need only compare the growth rate with other relevant timescales. 

If $\varphi_\text{min} > 0$, then classically the universe lasts forever, and one simply needs to wait long enough for the perturbations to grown large enough to form stars, \etc\ If there is negative vacuum energy then there exists a quantum instability to the nucleation of vacuum energy-dominated bubbles, which puts a lower bound on the vacuum energy.

Alternatively, if $\varphi_\text{min} \leq 0$, then the universe exists only for a finite time $t$, and so one needs to ensure that the perturbations have grown sufficiently by the time the universe attains its maximum size. This ends up putting a lower bound on the co-moving matter density, and an upper bound on $|\varphi_\text{min}|$.

\subsection{Other solutions to the gravity problem}

In the previous subsections, I have presented a purely scalar theory of gravity which may allow life in $2+1$ dimensions; this is not intended as a complete theory, but more as a proof-of-principle, and now I will briefly discuss a few other alternatives. 

One could make the whole metric dynamical (rather than just the conformal factor) to end up with a scalar-tensor theory. See \cite{Barrow:1986} for work studying certain aspects of three-dimensional scalar-tensor theory. By making the Planck mass for the tensor fluctuations large enough, one would expect to retain the desirable features of the point-source solution (namely stable orbits) while forcing any changes to occur very close to the source (and one might even hope to achieve the introduction of event horizons to clothe the singularity). The changes to the cosmological solution are harder to predict.

Note also that vacuum polarisation may give rise to an inverse square law (in the weak-field limit) for a point source in $2+1$ dimensional general relativity \cite{Soleng:1993yh}.

On the other hand, one could imagine a braneworld, in which matter is confined to a 2-brane, while gravity may propagate in the four-dimensional bulk. Usually in a braneworld scenario, one must work to hide the higher-dimensional nature of gravity, by warping the bulk \cite{Randall:1999vf}, making it finite \cite{ArkaniHamed:1998rs}, or giving a mass to the graviton \cite{Kaloper:2018dqn}, yet in this case the fact that gravity would behave in a higher-dimensional manner (namely, four dimensionally) is actually a desirable feature.

It is worth noting that where comparable observations can be made, all of these theories may depart from what it observed in our universe, however, for the question of whether life can exist, it does not seem necessary to fully reproduce gravity as in our universe (merely certain features, such as the existence of stable orbits).

\section{Biological Neural Networks} \label{sec-neural networks}

Driven by the study of complex networks in general (see \cite{Albert:2002, Newman:2003} for reviews), the study of biological neural networks is a flourishing field (see \cite{Bullmore:2009} for a review) and represents the analysis of both structural and functional networks, in creatures ranging from the nematode \textit{C. elegans}, whose entire network of neurons has been mapped \cite{White:1986}, to mammals such as cats \cite{Scannell:1999} and macaques \cite{Felleman:1991}, to humans themselves \cite{Sporns:2005}, though in the latter cases the data are not close to the resolution of individual neurons.

One key property that neural networks seem to have (along with many other real world networks) is that they are \emph{small-world} \cite{Watts:1998, Humphries:2006}, which means that they simultaneously exhibit both the high clustering of regular lattices and the small average path length of random graphs. It has been conjectured that the small-world property may be evolutionarily selected, due to providing both robustness and efficiency \cite{Latora:2001}; see \cite{Bassett:2017} for a recent review on small-world neural networks.

However it has recently been realised that this on its own may not be enough to characterise real world neural networks, and that in fact departures from pure small-world behaviour may be necessary \cite{Hilgetag:2015}. It would seem that another hallmark of brain networks is that they are critical, being poised between active and quiescent phases in which the neurons are either all firing or all dormant \cite{Petermann15921, Kitzbichler:2009} (although the issue of criticality is not entirely settled \cite{Beggs:2012}). It is conjectured that this is required for effective processing of stimuli \cite{Woodrow:2013}, since otherwise the firing of a neuron due to an external stimulus would either be swamped in noise (due to previous firings) or not lead to a response in the brain. Criticality generally requires tuning; however, this can be ameliorated by a \emph{Griffiths phase} \cite{Griffiths:1969}, a `stretching' of the critical point into a region (see \cite{Munoz:2010} and references therein). One property which seems important for the existence of a Griffiths phase is hierarchical modularity \cite{Moretti:2013, Odor:2015}, \ie the network is constructed from smaller sub-networks within which the connectivity is much higher than with the other modules.

The question I wish to ask is then: do there exist families of planar graphs that exhibit the properties which have so far been deemed important in real-world brain networks, namely, small-worldness, hierarchical modularity, and the existence of Griffiths phases?

\subsection{Classes of random planar graphs}

\subsubsection{(Uniform) Random planar graphs}

Defining a random planar graph (RPG) to be one drawn uniformly at random from the set of all planar graphs of a given number of (labelled) vertices, there is a simple Markov chain algorithm to generate one \cite{Denise:1996}. This, however, is rather slow, but faster algorithms have been developed and presented in \cite{Bodirsky:2007} and \cite{Fusy:2009};\footnote{See \cite{Meinert:2011} for a comparison of different generators random planar graphs.} an implementation of this last one can be found at \cite{Fusy:implementation} which will generate random planar graphs with approximately $10^3$, $10^4$, and $10^5$ vertices, and in what follows I will use a set of 20 graphs generated for each of these sizes.
See \cite{Gimenez:2005} for more information on the properties of random planar graphs.

These random planar graphs have no structure (beyond being planar), and so I will now describe two classes of planar graphs whose construction inherently introduces a certain amount of hierarchical structure and modularity (which seems to be important for the existence of a Griffiths phase).

\subsubsection{Self-similar (planar) graphs}

Start with an RPG and for each of its vertices, with probability $p$ identify that vertex with the vertex of another RPG, then repeat this procedure for all of the new sub-graphs that are introduced. In this way one generates a planar graph with a self-similar structure. If the graphs at each level have $N$ vertices, then after performing this procedure $n$ times, the expected number of vertices of the resulting graph is 
\begin{equation}
V_n = (1-p) N \left( 1 + \cdots + (pN)^{n-1} \right) + (pN)^n N = \frac{N}{pN - 1} \left( (pN)^{n+1} - 1 - p \left( (pN)^n - 1 \right) \right)
\end{equation}
and if the expected number of edges in an RPG is $\gamma_N N$ ($\gamma_N \approx 2.2$ for large $N$) then the expected average degree of our self-similar graph is
\begin{equation}
\langle k \rangle = 2 \frac{E_n}{V_n} = 2\gamma_N \frac{ 1 + \cdots + (pN)^n}{(1-p) \left( 1 + \cdots + (pN)^{n-1} \right) + (pN)^n} = 2 \gamma_N \frac{(pN)^{n+1} - 1}{(pN)^{n+1} - 1 - p \left( (pN)^n - 1 \right)}.
\end{equation}

In what follows I will use a set of graphs generated with $N = 10$, $p = 1$, and $n = 3,4,5$ (so their sizes are $10^3, 10^4, 10^5$---the same as the RPGs).

\subsubsection{Cycle-based (planar) graphs} \label{sec-cycle based}

\begin{figure}[tp]
\centering

\includegraphics[width=\textwidth]{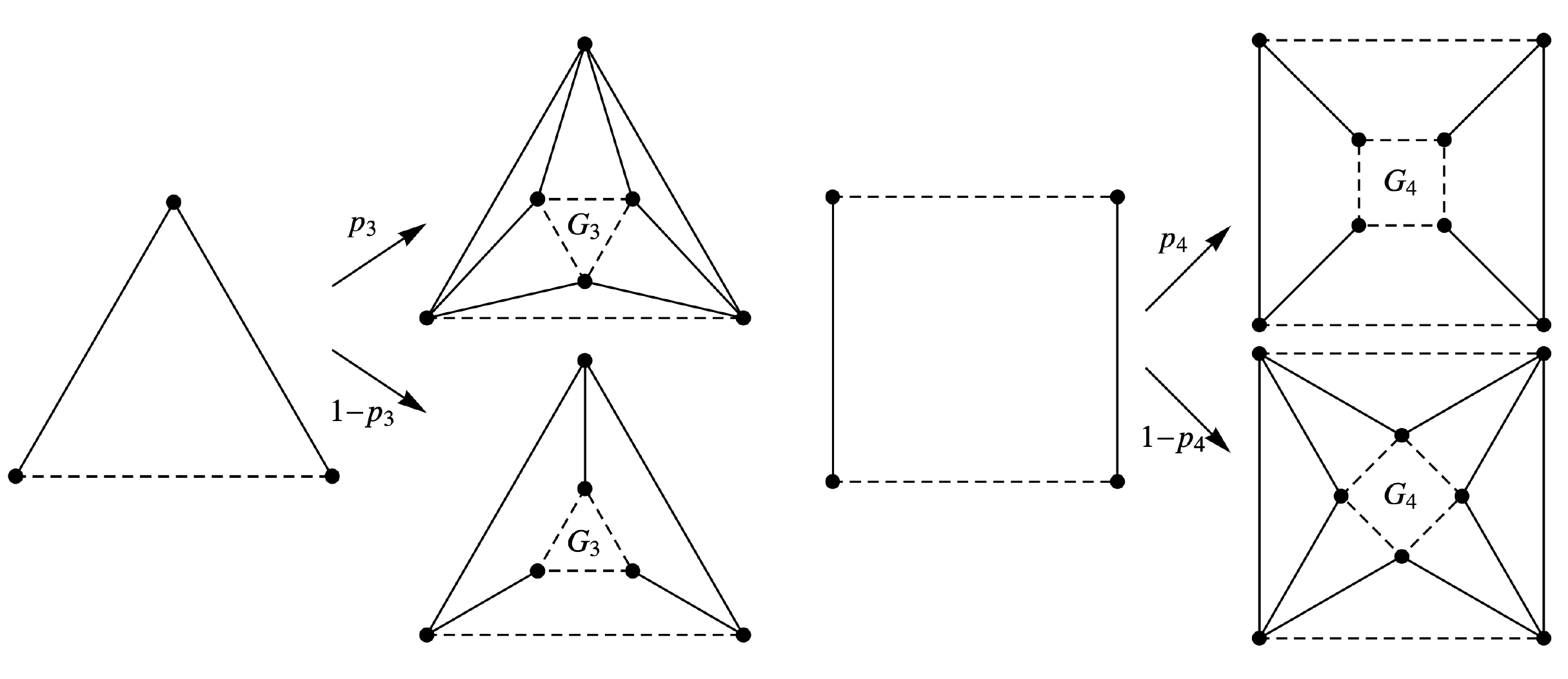}
\caption{The procedural generation of cycle-based random planar graphs. Dashed lines indicate edges which may or may not exist, depending on the edges of $G_3$ and $G_4$.} \label{fig-self-similar generation}
\end{figure}

Inspired by \cite{Zhang:2007, Chen:2008}, I will now define a class of random planar graphs whose generation is based on cycles. The starting point is a cycle graph of order three or four, and the process of generation follows thus (and is depicted in figure \ref{fig-self-similar generation}): given a specific set of three nodes, a planar graph $G_3$ is placed inside, and with probability $p_3$ three vertices of $G_3$ adjacent to the outside face are each connected to two of the vertices of the original set, creating six new graphs of order three, called \emph{active} three-graphs, which can then go on to spawn $G_3$'s of their own; alternatively, with complementary probability $1-p_3$, the three outerplanar vertices of $G_3$ may be each connected to just one vertex, which creates three active four-graphs; given a four-graph, a planar graph $G_4$ is placed inside, and with probability $p_4$ four outerplanar vertices are each connected to one of the vertices of the original graph, creating four active four-graphs; with complementary probability $1-p_4$ these vertices are each connected to two vertices of the original graph, creating eight active three-graphs. Thus, at each stage, active three-graphs and four-graphs generate new active three-graphs and four-graphs, and in this way the whole graph is procedurally generated. The only requirements on $G_3$ (resp. $G_4$) (which may of course be drawn from some random graph distribution) are that they are planar and have at least three (resp. four) vertices which may be drawn adjacent to the outside face.

Given the above rules, the expected number of active three/four-graphs at stage $n$ is
\begin{align}
N(3)_n &= 6 p_3 N(3)_{n-1} + 8 (1-p_4) N(4)_{n-1} \\
N(4)_n &= 3 (1-p_3) N(3)_{n-1} + 4 p_4 N(4)_{n-1},
\end{align}
where $(N(3)_0, N(4)_0 ) = (1,0) \text{ or } (0,1)$ depending on whether a three or four cycle is taken as the initial seed. It is simple to check that if one requires the expected number of active three-graphs to be in some fixed ratio $r$ to the expected number of active four-graphs, then one must pick $p_3 = \frac{r}{2+r}$ and $p_4 = \frac{2}{2+r}$.

The expected number of new vertices and edges added at stage $n$ is
\begin{align}
\Delta V_n &= N(3)_{n-1} V(G_3) + N(4)_{n-1} V(G_4) \\
\Delta E_n &= N(3)_{n-1} \left( p_3 \left(6 + E(G_3) \right) + (1-p_3) \left( 3 + E(G_3) \right) \right) \nonumber \\
&+ N(4)_{n-1} \left( p_4 \left(4 + E(G_4) \right) + (1-p_4) \left( 8 + E(G_4) \right) \right).
\end{align}
Hence if the expected number of active three-graphs is in some fixed ratio $r$ to the expected number of active four-graphs, then the average degree of the whole graph is approximately given by
\begin{equation}
\langle k \rangle \approx 2 \frac{\Delta E_n}{\Delta V_n} = 2 \frac{2(1+r)(4+3r) + (2+r)(E(G_4) + r E(G_3))}{(2+r) (V(G_4) + r V(G_3))} \leq 2 \left( \frac{2+2r}{2+r} + \frac{E(G_4) + r E(G_3)}{4+3r} \right),
\end{equation}
where the inequality arises from the fact that $V(G_3) \geq 3$ and $V(G_4) \geq 4$.

In what follows I will use a set of graphs generated with $p_3 = \frac{1}{3}$ and $p_4 = \frac{2}{3}$ (so that the ratio of three- to four-cycles is one), with equal probabilities for starting with a three- or a four-cycle, with three different classes of $G_3$ and $G_4$, as described below.
\begin{itemize}
\item $G_3$ and $G_4$ are the empty graphs on three and four vertices (\ie isolated vertices with no edges connecting them); this leads to $\langle k \rangle = \frac{8}{3} \approx 2.67$.
\item $G_3$ and $G_4$ are chosen randomly from the sets of labelled outerplanar\footnote{An outerplanar graph is a planar graph with the additional property that it can be drawn so that all of the vertices are adjacent to the outside face; this is done to ensure that the resulting cycle-based graph is planar.} graphs on three and four vertices; this leads to $\langle k \rangle = \frac{27}{7} \approx 3.86$.
\item $G_3$ and $G_4$ are each randomly chosen from the labelled outerplanar graphs (on three and four vertices) with the largest number of edges; this leads to $\langle k \rangle = \frac{104}{21} \approx 4.95$.
\end{itemize}
This leads to the expected number of vertices in the whole graph being approximately $V_n \approx \frac{21}{22} \left( \frac{14}{3} \right)^n$, and graphs are generated for $n = 5, \ldots, 9$, which is $N \approx 2 \times 10^3, 10^4, 5 \times 10^4, 2 \times 10^5, 10^6$.

\subsection{Topological and other properties of these classes of graphs}

In this section I will consider various topological properties of the previously described classes of graphs and compare them to one another as well as to (Erd\"os-R\'enyi) random graphs (which are not necessarily planar). 

\subsubsection{Small-worldness}

The average path length of a graph, $l$, is the shortest distance between two vertices, averaged over all pairs of vertices. For a random graph, with fixed average degree, this increases logarithmically with the number of vertices, 
and one has \cite{Albert:2002}
\begin{equation}
l_{ER} \sim \frac{\ln N}{\ln \langle k \rangle}.
\end{equation}
Whereas for a regular lattice, this would increase as a power law in $N$.

The local clustering coefficient of a vertex in a graph is the number of neighbours of the vertex which are themselves connected, divided by the total number of possible connections between the neighbours ($\frac{1}{2} k (k-1)$ if the vertex has degree $k$). This is then averaged over all the vertices to get the clustering coefficient of the graph. For a random graph, this is just the probability that any given possible edge is present and hence
\begin{equation}
C_{ER} = \frac{\langle k \rangle}{N},
\end{equation}
\ie for fixed average degree, it decreases as the graph size increases, whereas other graphs may exhibit much larger clustering.

A small-world graph \cite{Watts:1998} is one which exhibits the short average path length of random graphs but also has a much higher clustering coefficient; this can be recorded in the small-world coefficient
\begin{equation}
\sigma = \frac{\frac{C}{C_{ER}}}{\frac{l}{l_{ER}}}, \label{small-world coefficient}
\end{equation}
and if $\sigma > 1$, the graph is said to be small-world.

The average path length is found to increase logarithmically for the cycle-based and self-similar graphs, whereas for random planar graphs it increases as a power law with exponent $\approx 0.29$.

Meanwhile, the clustering coefficient is approximately constant for all the families of graphs as the size changes. If one imagines a planar graph growing via some process, then as new vertices and edges are added, the planarity constraint forces the new connections to more likely be between `nearby' vertices (in the sense that they share a neighbour) than between vertices separated by a larger distance.

\begin{figure}[tp]
\centering
\includegraphics[width=\textwidth]{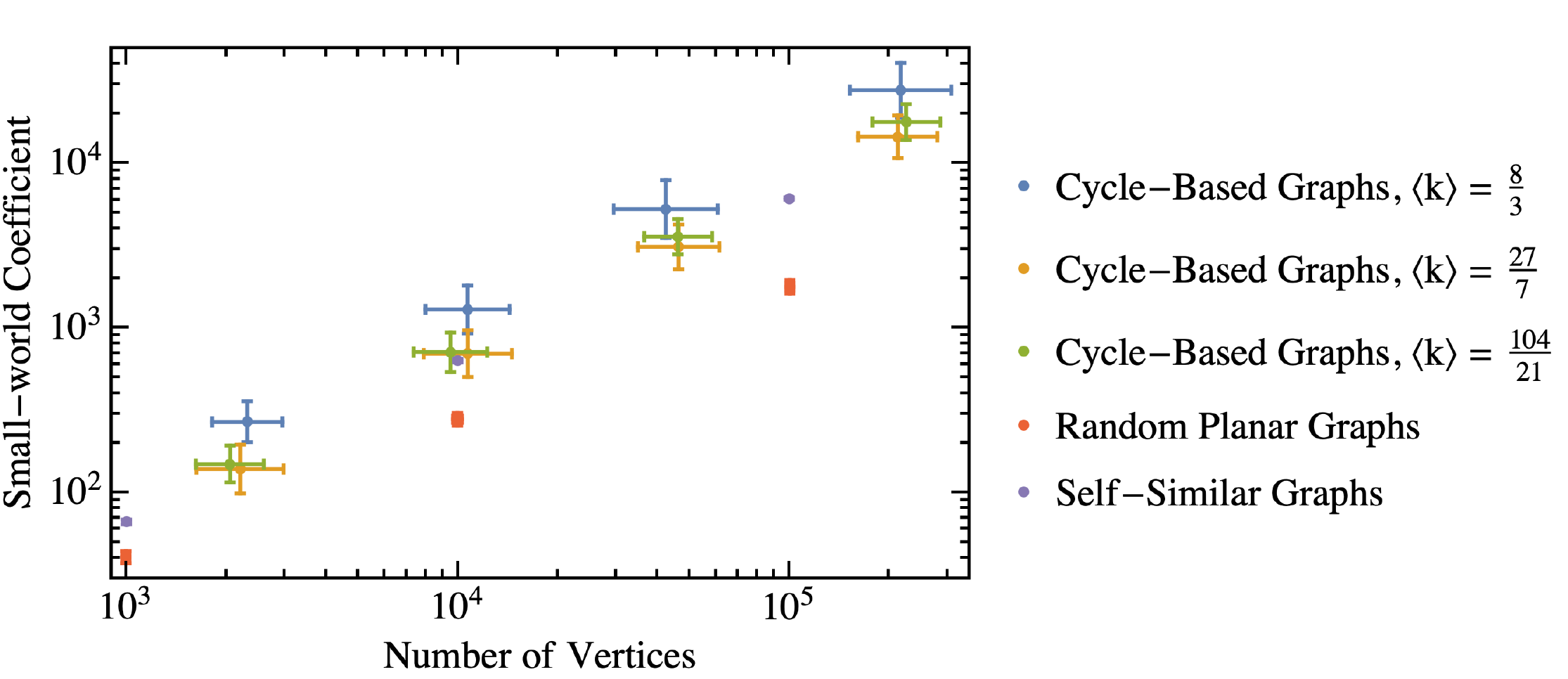}
\caption{The small-world coefficient, $\sigma = \frac{C / C_{ER}}{l / l_{ER}}$ for different families of random planar graphs. The cycle based graphs show the largest small-world coefficient (but also the largest scatter), whereas the RPGs show the smallest (though it is still significantly larger than one).}\label{fig-small-world}
\end{figure}

These facts combine to mean that all the families of random planar graphs are small-world, with the small-world coefficient increasing with the size of the graph, as shown in figure \ref{fig-small-world}.

\subsubsection{Topological Dimension}

The topological dimension of a graph is defined by the behaviour of $N(r)$, the number of vertices within a distance of $r$ a given node, averaged over the whole graph.\footnote{Given the size of the graphs under consideration, I will not actually average over every vertex, instead taking a random sample of $100$ vertices.} If $N(r) \sim r^D$, then the graph is said to have topological dimension $D$, whereas if it increases more quickly (exponentially) the graph has infinite topological dimension. It has been argued that biological neural networks possess finite topological dimension \cite{Hilgetag:2015}.

\begin{figure}[tp]
\centering
\includegraphics[width=\textwidth]{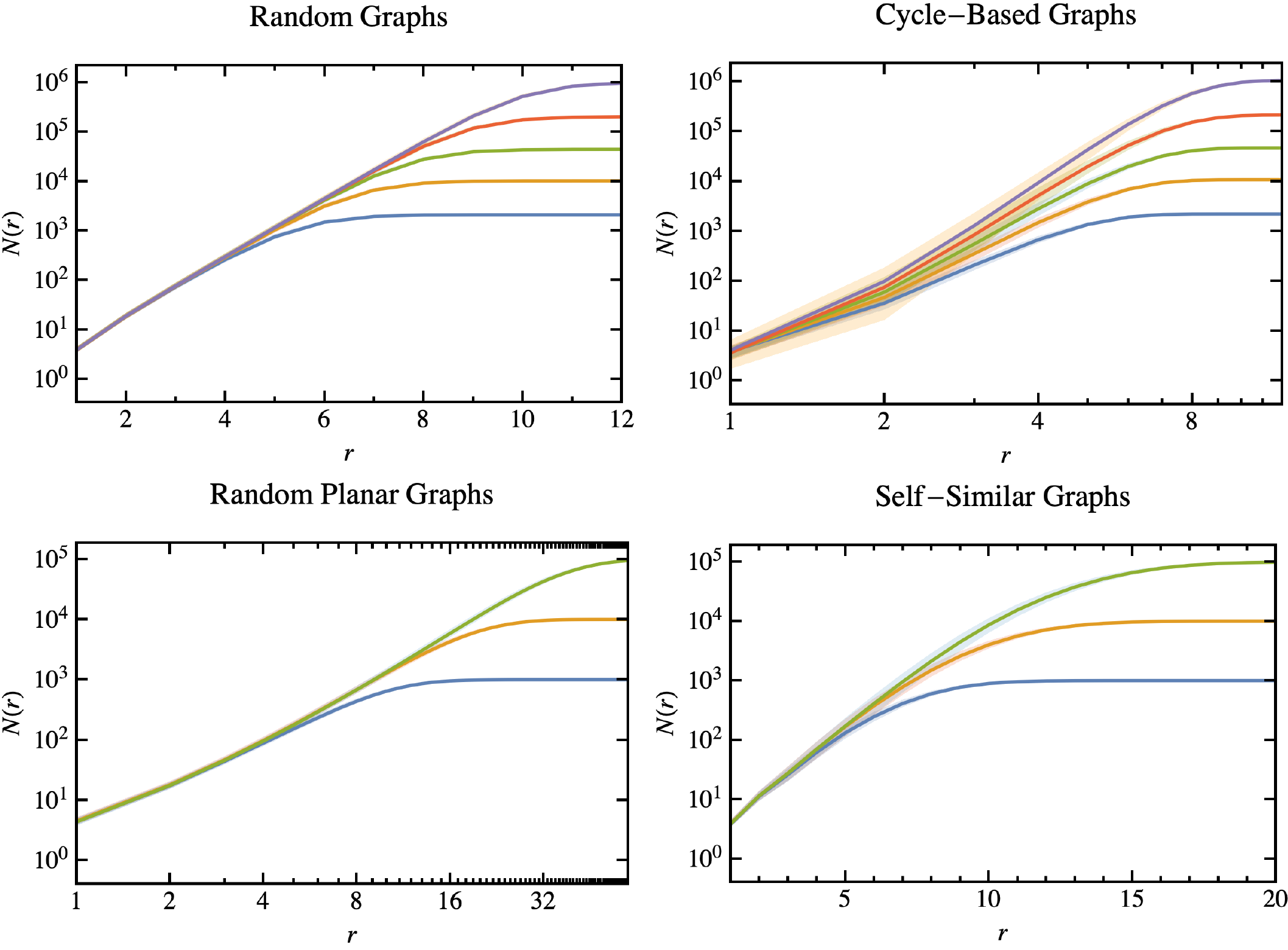}
\caption{The number of vertices within a given distance, averaged over 100 starting vertices, for four different families of graphs, for various sizes of graph; the light coloured bands indicate the uncertainty. The cycle-based and random planar graphs exhibit finite topological dimension, $N \sim r^D$ (though in the former case $D$ increases with graph size), whereas the random graphs and self-similar planar graphs exhibit infinite topological dimension, $N \sim \exp(r)$.}\label{fig-topological dimension}
\end{figure}

Plots of $N(r)$ can be found in figure \ref{fig-topological dimension}. Both random graphs and the self-similar planar graphs have infinite topological dimension ($N(r) \sim \exp(r)$) whereas random planar graphs seem to have finite topological dimension $D \approx 3$. The cycle-based planar graphs have larger, but still finite, topological dimension, however it increases logarithmically with the number of vertices from $D \approx 4$ at $N \approx 2 \times 10^3$ to $D \approx 6.5$ when $N \approx 10^6$, which leads one to suspect that in the infinite-size limit these graphs would actually have infinite topological dimension. This is independent of the average degree of the graph.

\subsubsection{Global reaching centrality}

Another feature of neural networks which goes beyond small-worldness is that they possess a hierarchical and modular construction \cite{Hilgetag:2015}.

A proposed measure of hierarchy in complex networks is the global reaching centrality \cite{Mones:2012}. For undirected graphs this is the average of the deviation of the harmonic centrality of each vertex from the maximum. The harmonic centrality of a vertex is the average of the reciprocals of the distances to all other vertices and thus is large if the vertex is close to many other vertices, and thus the global reaching centrality is large for graphs which have a few `hub' vertices.

\begin{figure}[tp]
\centering
\includegraphics[width=\textwidth]{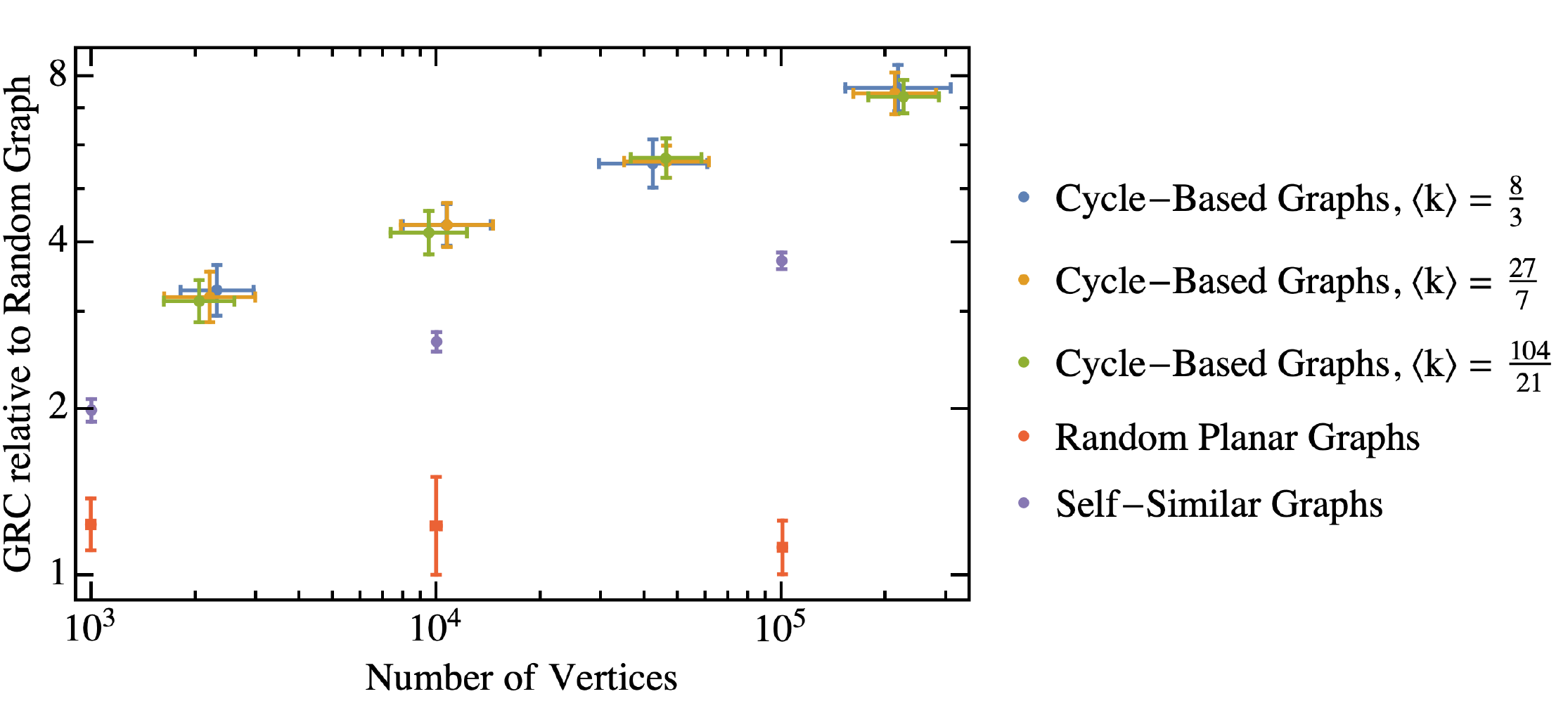}
\caption{The ratio of the global reaching centrality (GRC), a measure of a graph's hierarchical structure, of different families of random planar graphs to that of random graphs. The cycle based graphs show the largest GRC, whilst the RPGs show the smallest.}\label{fig-GRC}
\end{figure}

Numerical investigations show that for random graphs $GRC \sim N^{-1/4}$, with a weak dependence on the average degree of the graph (which is smaller than the intrinsic scatter). In figure \ref{fig-GRC} is plotted the ratio of the global reaching centrality for each of the families of graphs to that of equivalent random graphs; one can see that the GRC of random planar graphs is similar to that of random graphs, whereas the self-similar and cycle-based graphs posses a GRC which is larger than random graphs by a factor which grows with size. This is unsurprising given that their construction is explicitly hierarchical.

\subsubsection{Degree distribution} \label{sec-degree distribution}

An important property of a graph is the distribution of the degrees of its vertices. At this point it is worth mentioning that the average degree is one way in which planar graphs can never resemble biological neural networks, for these typically have very large average degree---for the human connectome, it is $\langle k \rangle \sim 10^3$, while even \textit{C. elegans} has $\langle k \rangle \approx 23$---whereas planar graphs have $\langle k \rangle < 6$ (when embedded in a space of genus zero).

In similarity with random graphs, the degree distribution of RPGs and the self-similar graphs has an exponential tail at large degree, and for these graphs the expected maximum degree thus increases logarithmically with graph size. 

On the other hand, the cycle-based graphs show a quasi-scale-free degree distribution, with an approximately power law tail (due to the nature of their construction, there are some weak oscillatory features in it) with $P(k) \sim k^{-3}$. As a result, the expected maximum degree of these graphs is much larger, and increases as a power law in the graph size. Therefore although the average degree is much smaller than that found in real life neural networks, a fraction of the vertices in fact do have comparably large degrees.

\subsubsection{Inverse Participation Ratio} \label{sec-IPR}

The inverse participation ratio, defined by
\begin{equation}
IPR = \sum_{i=0}^N \left( v^{(1)}_i \right)^4,
\end{equation}
where $v^{(1)}$ is the principal eigenvector of the adjacency matrix of a graph, measures how localised this eigenvector is. If it is de-localised then $v_i \sim N^{-1/2}$ and hence $IPR \sim N^{-1}$, whereas if $v^{(1)}$ takes non-zero values only on a subset of the vertices, \ie it is localised, then the inverse participation ratio may be much larger. Thus it is a diagnostic of localisation, and hence the `rare regions' which are important for the existence of a Griffiths phase.

\begin{figure}[tp]
\centering
\includegraphics[width=\textwidth]{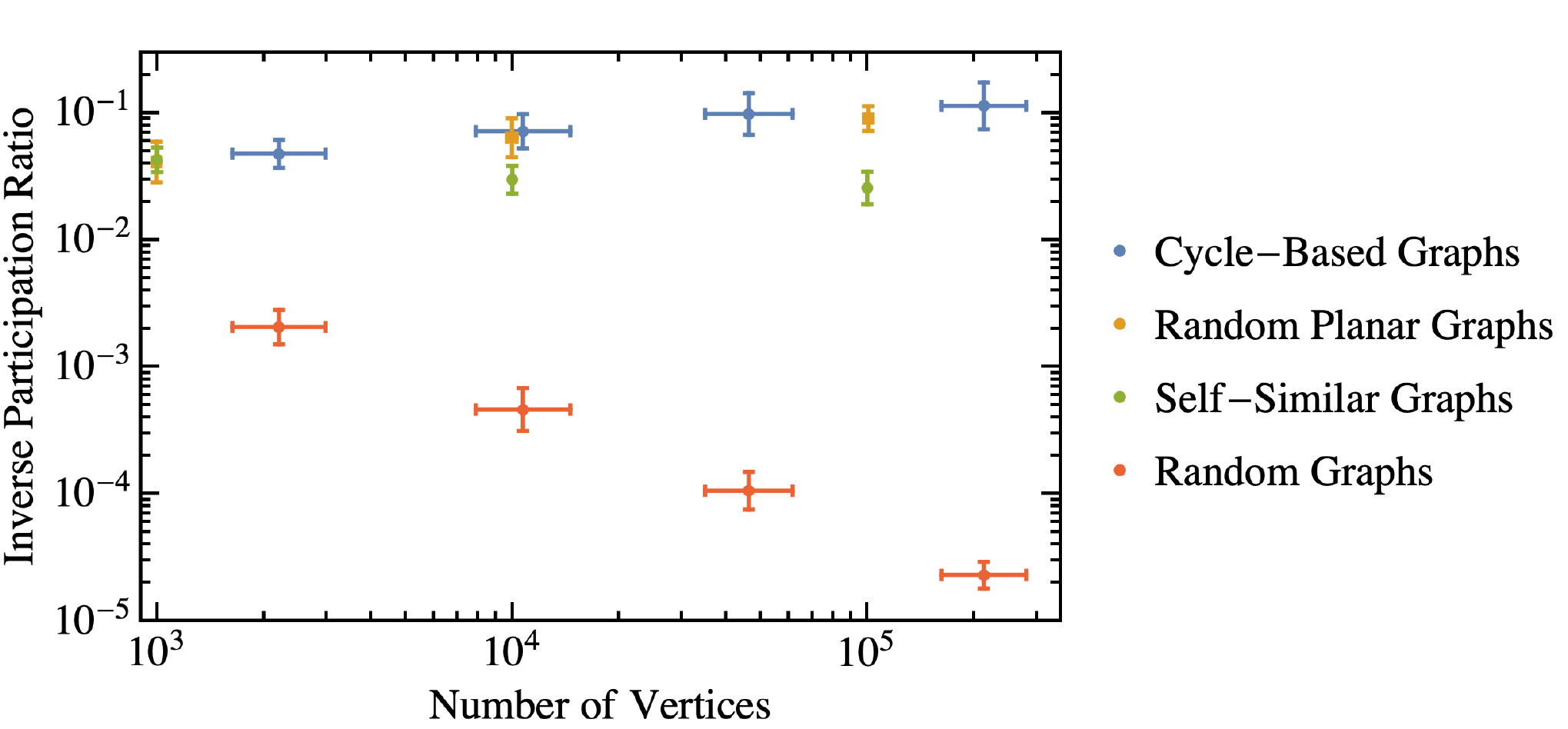}
\caption{The inverse participation ratio for different families of graphs (for the cycle-based and random graphs the average degree is $\langle k \rangle = \frac{27}{7}$). The random graphs exhibit no localisation: $IPR \sim N^{-1}$, whereas all the other families of graphs exhibit some degree of localisation, though it is weakest for the self-similar graphs.}\label{fig-IPR}
\end{figure}

Its behaviour is plotted in figure \ref{fig-IPR}. One indeed sees delocalised ($IPR \sim N^{-1}$) behaviour for random graphs; however, all of the other families of graphs show some degree of localisation. (Yet as we will see later, this on its own is not sufficient for the existence of a Griffiths phase.)

\subsection{Dynamics}

Let us now turn to considering stochastic processes on the graphs described above, in order to investigate the existence of Griffiths phases. In particular we will be considering processes in which the vertices of the graph can be either active or inactive, and study how the total activation density behaves in time, starting from an initial state which is fully active. Before presenting the results, let us first briefly show how the presence of disorder can lead to a Griffiths phase; this example follows \cite{Vojta:2006}. 

Consider a process which exhibits a phase transition as the `spreading rate' $\lambda$ is varied; when $\lambda$ is below some critical value $\lambda_c$, the system is in the inactive phase and the activation density $\rho$ will decay exponentially; alternatively, when $\lambda > \lambda_c$, the density will approach some finite ($\lambda$-dependent) value at late times; these two phases are separated by a critical point for which the density decays as a power law.

Now, rather than a uniform spreading rate, consider one which is site-dependent, with the spreading rate at a particular site taking value $\lambda_1$ with probability $p$, and $\lambda_2 > \lambda_1$ with probability $1-p$. Clearly, if $\lambda_2 < \lambda_c$, then the system will be in the inactive phase, and vice versa for $\lambda_1 > \lambda_c$, but when $\lambda_1 < \lambda_c < \lambda_2$, the situation is more complicated. Regions in which the spreading rate is mostly $\lambda_1$ will quickly become inactive whereas the rare regions in which the spreading rate is mostly $\lambda_2$ will stay active for much longer, dominating the late-time activation density. 

Their finite size, and the presence of regions in the inactive phase, means that they cannot sustain infinitely long-lived activation, however, it does require a coordinated fluctuation to deactivate them, and thus they are exponentially long-lived, $\tau \sim \exp(a N)$, for a region of size $N$, where $a$ is some constant which depends on $\lambda_2$ (and which goes to zero when $\lambda_2$ goes to $\lambda_c$). On the other hand such regions are exponentially rare. The late-time activation density is thus
\begin{equation}
\rho(t) \sim \int dN N (1-p)^N \exp\left( - \frac{t}{\tau(N)} \right) = \int dN \exp \left( -t \exp(-a N) + N \ln (1-p) + \ln N \right)
\end{equation}

Solving this in the saddle-point approximation, one finds
\begin{equation}
\rho(t) \sim t^{\ln(1-p) / a(\lambda_2)}.
\end{equation}
That is, the activation density decreases as a power law, with an index which varies continuously as the spreading rate $\lambda_2$ is varied.

In the case being considered here, the spreading rate itself is actually uniform, however the structure of the graph (potentially) generates the required disorder (\eg through disorder in vertex degree).\footnote{Of course, the analysis in the given example tacitly assumed that the disorder is spatially uncorrelated, whereas disorder due to vertex degrees may be correlated if, for example, the graph exhibits preferential attachment of high degree nodes to another.}

\subsubsection{Results}

Let us first consider the stochastic process whose evolution equation is
\begin{equation}
\dot{\rho}_i = - \mu \rho_i + \lambda (1-\rho_i) \sum_j A_{ij} \rho_j, \label{SIS}
\end{equation}
where $\rho_i \in \{0, 1\}$ is the state of neuron $i$, and $A_{ij}$ is the adjacency matrix of the graph. This equation describes a scenario in which active neurons deactivate with density $\mu$, and quiescent neurons are switched on by their active neighbours with rate $\lambda$; the dynamics is just controlled by the ratio of $\lambda$ to $\mu$ and so, without loss of generality, one may set $\mu$ to unity.

\begin{figure}[tp]
\centering
\includegraphics[width=\textwidth]{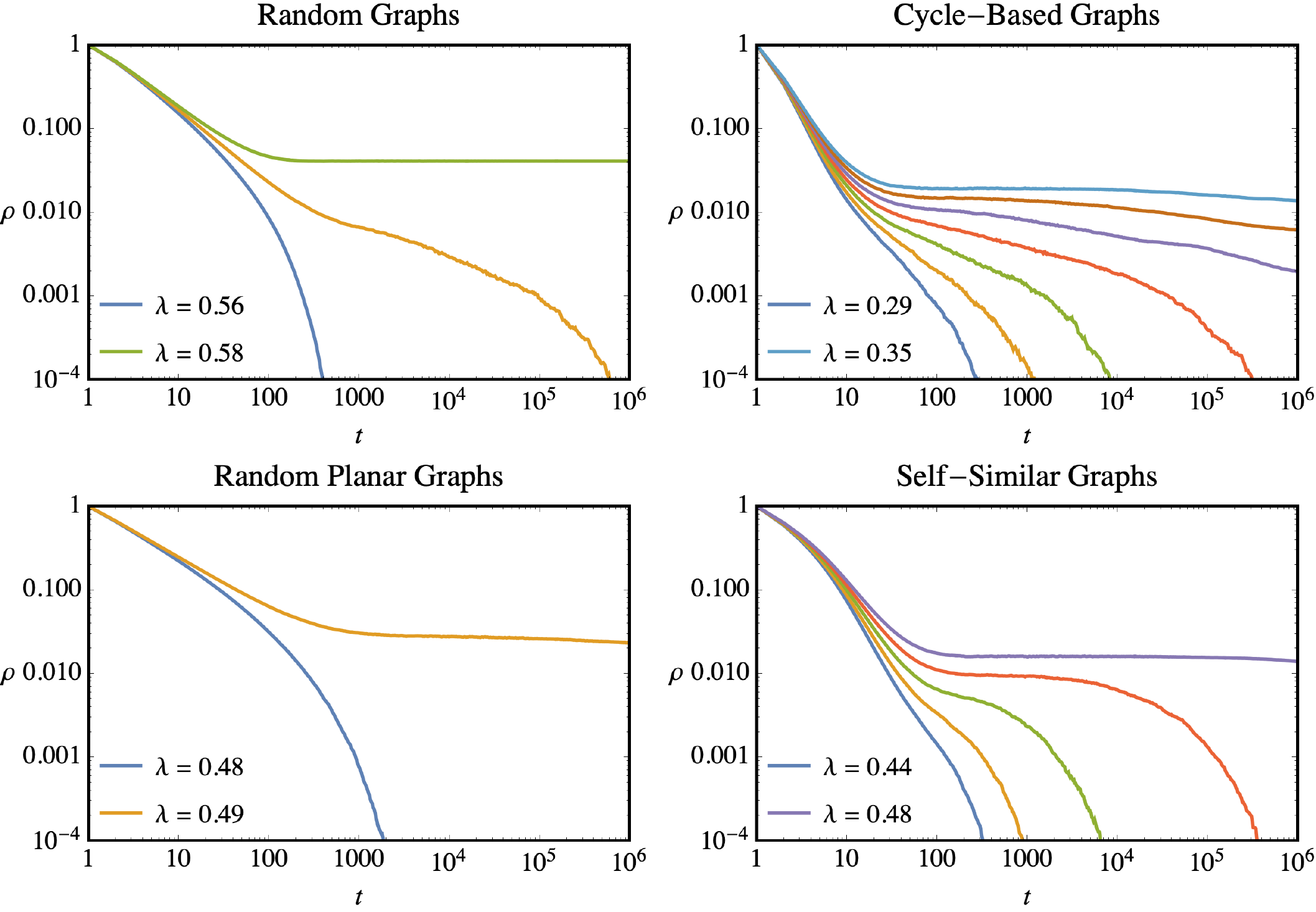}
\caption{The results of running the stochastic process described by \eqref{SIS} (with the deactivation rate, $\mu$, set to unity) on four families of graphs (with 20 graphs in each family, with the process run 50 times on each graph), starting from a fully active state, for various values of the spreading rate $\lambda$ (with spacing $\Delta \lambda = 0.01$). The random graphs and cycle-based graphs have $\langle N \rangle \approx 2.2 \times 10^5$ and $\langle k \rangle \approx 2.7$, while the random planar graphs and self-similar graphs have $\langle N \rangle \approx 10^5$ and $\langle k \rangle \approx 4.4$ and $\langle k \rangle \approx 3.9$ respectively.}\label{fig-SIS}
\end{figure}

The results of running this on the previously described families of random graphs are shown in figure \ref{fig-SIS}. Edr\"os-R\'enyi (non-planar) random graphs (top left) exhibit a relatively sharp transition from a quiescent phase to an active phase; random planar graphs (bottom left) also exhibit a sharp transition; the self-similar graphs (bottom right) show slightly unusual behaviour over a range of spreading rates, with the density almost settling down to a constant value (as in the active phase) before decaying exponentially (as in the quiescent phase), but there is no period of power law decay. Finally, the cycle-based graphs (top right) do seem to show evidence of a Griffiths phase, exhibiting power-law decay over a range of spreading rates. The results shown here are for $\langle k \rangle = \frac{8}{3}$, and a Griffiths phase is also observed for larger average degrees, albeit over a slightly reduced range in $\lambda$.

The long-time dynamics is governed by the principal eigenvector of the adjacency matrix and hence is directly related to the inverse participation ratio discussed in section \ref{sec-IPR}. And indeed one sees that the random graphs have both a delocalised prinicpal eigenvector and a sharp phase transition, and the cycle-based graphs show the opposite (on both counts). On the other hand, both the random planar graphs and the self-similar graphs exhibit some degree of localisation, but present no evidence of a Griffiths phase.

Finally, it should be noted that while for the random graphs and RPGs the observed behaviour is size-independent, for the self-similar and cycle-based graphs, the characteristic value of the spreading rate decreases as the graph size increases. This is because the behaviour in those cases is driven by the highly inhomogeneous degree distribution, which, due to the way in which these graphs are constructed, changes as the graph size increases (in particular the maximum degree increases more quickly).

\begin{figure}[tp]
\centering
\includegraphics[width=\textwidth]{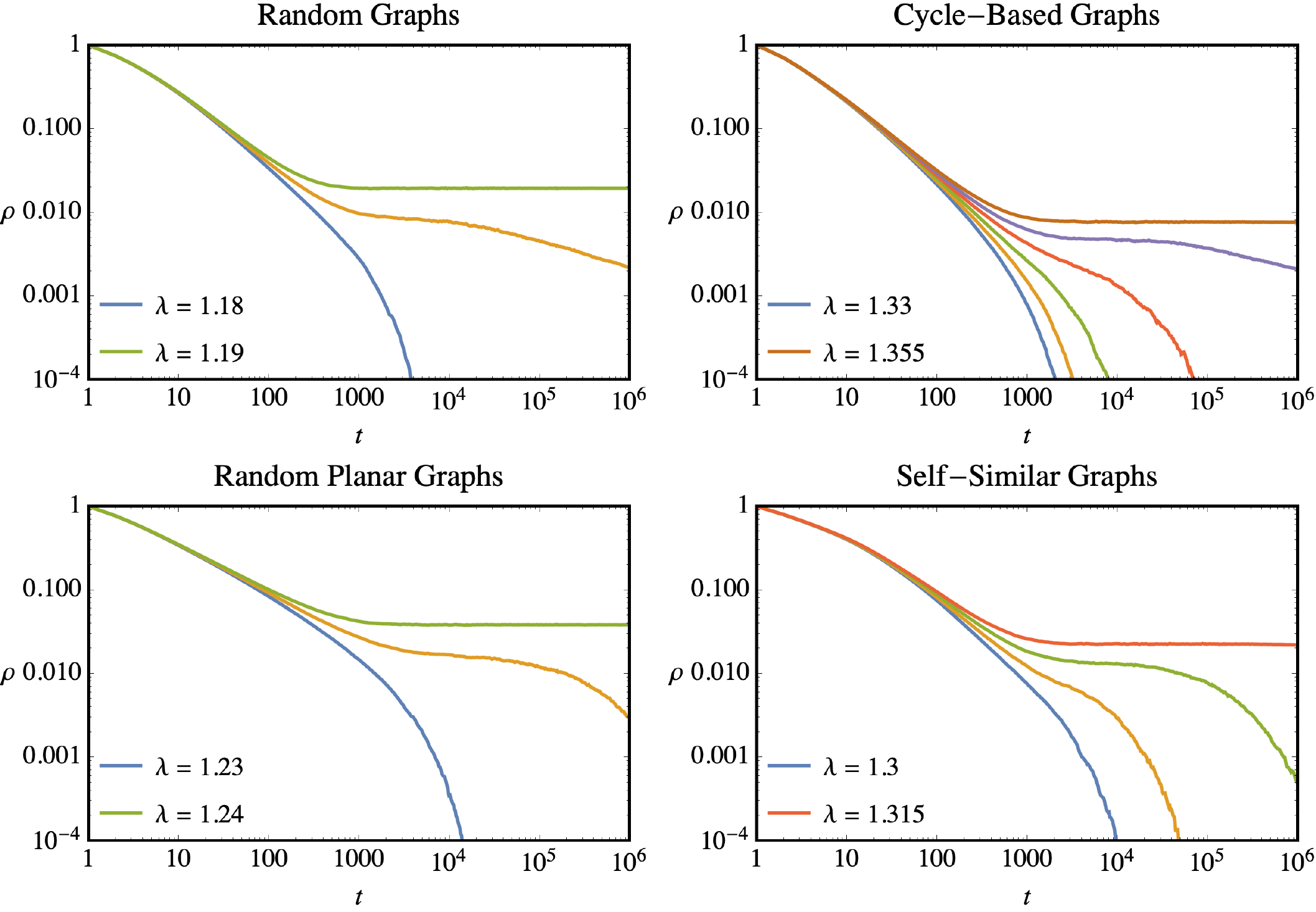}
\caption{The results of running the stochastic process described by \eqref{CP} on four families of graphs (with 20 graphs in each family, with the process run 50 times on each graph), starting from a fully active state, for various values of the spreading rate $\lambda$ (with spacing $\Delta \lambda = 0.005$). The random graphs and cycle-based graphs have $\langle N \rangle \approx 2.2 \times 10^5$ and $\langle k \rangle \approx 2.7$, while the random planar graphs and self-similar graphs have $\langle N \rangle \approx 10^5$ and $\langle k \rangle \approx 4.4$ and $\langle k \rangle \approx 3.9$ respectively.}\label{fig-CP}
\end{figure}

This last fact can be avoided by normalising the spreading rate by vertex degree, so that the evolution equation is 
\begin{equation}
\dot{\rho}_i = - \rho_i + \lambda (1-\rho_i) \sum_j \frac{A_{ij}}{\sum_k A_{jk}} \rho_j, \label{CP}
\end{equation}
where again the deactivation rate has been set to unity. The results for this can be found in figure \ref{fig-CP}, and one sees that the behaviour is qualitatively similar: the random graphs and random planar graphs show a sharp transition, the self-similar graphs do not show a period of power decay, and the cycle-based graphs show some evidence of power-law decays for a range of spreading rates although this is not as pronounced as in the previous case.

\section{Some Other Considerations concerning Physics in Three Dimensions} \label{sec-other things}

Finally, it behoves me also to briefly consider some other aspects of physics in three dimensions which may have a bearing on life; in particular, electromagnetism and the existence and structure of atoms and molecules.

\subsection{Maxwell's equations}

As in four dimensions, the electromagnetic field strength tensor is $F_{\mu\nu} = 2\partial_{[\mu} A_{\nu]}$, the electric field is $E_i = F_{0i}$, and the magnetic field $B = -\frac{1}{2} \epsilon_{0\mu\nu} F^{\mu\nu}$. The first thing to note is that while the electric field is still a vector, the magnetic field is now a (pseudo-)scalar. As in four dimensions, we can take the divergence of the field strength,
\begin{equation}
\partial_\mu F^{\mu\nu} = \mu_0 J^\nu,
\end{equation}
where $J^\mu = (\rho, \mathbf{j})$ is the current three-vector, to get the two inhomogeneous equations,
\begin{equation}
\nabla \cdot \mathbf{E} = \frac{\rho}{\epsilon_0}, \qquad \nabla \times B - \dot{\mathbf{E}} = \mu_0 \mathbf{j},
\end{equation}
where $(\nabla \times)_i = \epsilon_{ij} \nabla_j$. The first equation is identical in form to its four dimensional counterpart, while the second is identical to Amp\`ere's law in four dimensions, if the magnetic field is constrained to lie only in one direction and the electric field is constrained to be perpendicular to it. The Bianchi identity, $\partial_{[\rho} F_{\mu\nu]} = 0$, in three dimensions gives just one equation,
\begin{equation}
\dot{B} - \nabla \times \mathbf{E} = 0,
\end{equation}
which is identical (apart from a sign due to the curl definition) to Faraday's law in four dimensions with the same restrictions as mentioned above. There is no analogue of the magnetic version of Gauss' law.

These equations can of course also be derived simply by setting $E_z = B_x = B_y = 0$ in the four dimensional Maxwell equations.

It is easy to see that these equations still imply electromagnetic waves, with the vector electric field exciting the pseudo-scalar magnetic field and vice-versa.

A point electric charge will support an electric field which decays inversely with the distance from the charge, and hence the potential increases logarithmically, which means that an infinite amount of energy would be required to separate two charges. However, this undesirable result can be avoided by giving a small mass to the photon, which will then lead to a Yukawa-like suppression of the electric field ($E \propto H^{(1)}_1(imr) \sim e^{-mr}/\sqrt{mr}$ for large $r$), rendering the required energy finite. Since the gauge symmetry is broken by the mass term, there will also be an extra degree of freedom, however, it will not couple to conserved currents---as can be easily seen using the St\"ukelberg trick---and hence will not mediate an extra force.

\subsection{Atoms and molecules}

In his book \emph{What is Life?} \cite{WhatIsLife}, Schr\"odinger notes that quantum mechanics may be required for life. This is because hereditary information must be carried by something which is both small and stable. Molecules (and, in particular, complex DNA molecules, though Schr\"odinger did not know of them at the time) are a possible solution; however, classical physics cannot explain their stability---quantum mechanics is required for this. Therefore, the question is: are there any impediments in two spatial dimensions to the existence of stable (complex) molecules?

First of all, we should note that although more general statistics are possible (\viz anyons), in two spatial dimensions we can still have particles which satisfy Fermi-Dirac statistics and hence the Pauli exclusion principle. This means that atoms can form and we can build a periodic table of elements in the usual way, and in fact this was done by Dewdney and Lapidus in \cite{2dscitech, 2ndSymposiumon2dscitech}. Note that the resulting array of elements is somewhat simpler than in three dimensions, since the electron shells fill up more quickly,\footnote{Whereas in three dimensions a shell with principle and orbital angular momentum quantum numbers $n$ and $l$ can fit $2l+1$ electrons of a given spin, in two dimensions each shell can only fit two, or one for $l = 0$.} however, it is not completely devoid of structure.

When constructing molecules, the main difference introduced by the restriction to two dimensions is that molecules must be planar since the inter-atomic bonds cannot cross. Again this means that the possibilities are somewhat fewer (although the prevalence of chiral molecules is greater), though again not disastrously so, and, in fact, a plausible biochemistry formed of two dimensional molecules was developed in \cite{2ndSymposiumon2dscitech}.

\section{Conclusions}

In this paper, I have considered the two main arguments which are commonly presented against the possibility of complex life in $2 + 1$ dimensions: the absence of a local gravitational force in three-dimensional general relativity, and that the topological restrictions placed by requiring planarity are too severe to allow complex life.

The first can be avoided by changing the gravitational theory. As a proof-of-principle I have presented a purely scalar theory of gravity which allows stable orbits around point sources, and has a not-obviously-fatal (though unusual) cosmology; it could potentially be improved by making the whole metric dynamical. One could also imagine a brane-world scenario in which the massless graviton is \emph{not} localised to the brane, thus allowing two-dimensional life to enjoy fully four-dimensional gravity.

To deal with the second objection, I have turned to research that has been conducted on the properties of biological neural networks, and created a family of planar graphs (the `cycle-based' ones from section \ref{sec-cycle based}) which seem to exhibit many of the properties which have been conjectured to be important for complex brains. In particular, they are approximately `small-world,' they have a hierarchical and modular construction, and they show evidence of the stretching (in parameter space) of a critical point into a finite critical region for certain stochastic processes. It should be noted that while this is certainly suggestive of the possibility of complex brains in two dimensions, it is not conclusive, as it is likely that the properties described above are not sufficient on their own. Therefore more work is needed to compare the graphs presented here with real-life neural networks (and also to include edge weights and directions, which have been neglected here), as well as to consider further families of planar graphs, in order to arrive at a more complete understanding of the possibility of complex brains in two dimensions.

Overall it would seem that if one wishes to use anthropic reasoning to explain the observed dimensionality of space-time, then the possibility of life in $2 + 1$ dimensions requires further investigation. In particular, it would be interesting to determine if there might be other impediments to life which have so far been overlooked,\footnote{For example, it has been suggested that the fact that random walks almost surely return to their starting point in two spatial dimensions, whereas they do so with only finite probability in three spatial dimensions, might have relevance for decoherence and the emergence of a classical world \cite{AndysIdea}.} as well as to continue to search for non-anthropic explanations for the dimensionality of space-time.

\section*{Acknowledgements}

I would like to thank A. Albrecht and S. Carlip for useful comments and suggestions. This work was supported in part by DOE Grants DE-SC0009999 and DE-SC0019081.

\bibliographystyle{aps}
\bibliography{2D_brains}

\end{document}